\documentclass[a4paper,11pt]{article}
\usepackage{jheppub}
\usepackage[utf8]{inputenc}
\usepackage{url}
\usepackage{mathrsfs}  
\usepackage{pdflscape}

\usepackage{refcount}
\usepackage{booktabs}
\usepackage{todonotes}
\usepackage{enumitem}
\usepackage{adjustbox}
\usepackage{afterpage}
\usepackage{dsfont}
\usepackage{nameref}
\usepackage{dsfont}
\usepackage{makecell}

\usepackage{longtable}
\usepackage{float}

\makeatletter
\newcommand\level[1]{%
  \ifcase#1\relax\expandafter\chapter\or
    \expandafter\section\or
    \expandafter\subsection\or
    \expandafter\subsubsection\else
    \def\next{\@level{#1}}\expandafter\next
  \fi}
\newcommand{\@level}[1]{%
  \@startsection{level#1}
    {#1}
    {\z@}%
    {-3.25ex\@plus -1ex \@minus -.2ex}%
    {1.5ex \@plus .2ex}%
    {\normalfont\normalsize\bfseries}}

\newcounter{level4}[subsubsection]
\@namedef{thelevel4}{\thesubsubsection.\arabic{level4}}
\@namedef{level4mark}#1{}
\count@=4
\loop\ifnum\count@<100
  \begingroup\edef\x{\endgroup
    \noexpand\newcounter{level\number\numexpr\count@+1\relax}[level\number\count@]
    \noexpand\@namedef{thelevel\number\numexpr\count@+1\relax}{%
      \noexpand\@nameuse{thelevel\number\count@}.\noexpand\arabic{level\number\numexpr\count@+1\relax}}
    \noexpand\@namedef{level\number\numexpr\count@+1\relax mark}####1{}}
  \x
  \advance\count@\@ne
\repeat
\makeatother
\usepackage{tikz}
\usetikzlibrary{decorations.pathreplacing, positioning}

\usepackage[utf8]{inputenc} 
\usepackage{amsmath}
\usepackage{amsthm}
\usepackage{amsfonts,mathrsfs} 
\usepackage{xcolor}
\usepackage[english]{babel} 
\usepackage[autostyle]{csquotes}
\usepackage{mathtools}

\newlist{inparaenum}{enumerate}{3}
\setlist[inparaenum,1]{label=\arabic*.}
\setlist[inparaenum,2]{label=\emph{\alph*})}
\setlist[inparaenum,3]{label=\emph{\roman*})}

\numberwithin{equation}{section}

\makeatletter
\newcommand\footnoteref[1]{\protected@xdef\@thefnmark{\ref{#1}}\@footnotemark}
\makeatother

\preprint{Imperial/TP/25/AH/07}

\title{Higgs branch of 5d $\mathcal{N}=1$ symplectic gauge theories and dressed instanton operators}

\author{Amihay Hanany}
\author{and Elias Van den Driessche}
\affiliation{Theoretical Physics Group, Blackett Laboratory, Imperial College London, Prince Consort Road
London, SW7 2AZ, UK}
\emailAdd{a.hanany@imperial.ac.uk}
\emailAdd{e.van-den-driessche24@imperial.ac.uk}

\begin{document}

\abstract{We expand in instanton charge sectors the representation content of the infinite coupling chiral ring of the Higgs branch of 5d $\mathcal{N}=1$ $Sp(k)$ theories with $N_f$ flavours. The entire chiral ring can be expressed as the product of bare instantons, one for each topological sector, times a common dressing factor depending on the mesons and the instanton-anti instanton bound state. The dressing factor, which is independent of the instanton number, encodes the chiral ring of the theory at finite coupling with one additional colour.}

\maketitle

\section{Introduction}
As discovered in the late 90s by Seiberg and collaborators \cite{seiberg1996five,intriligator1997five,morrison1997extremal}, in many cases 5d $\mathcal{N}=1$ gauge theories admit a 5d strongly coupled fixed point with enhanced global symmetry, despite being power-counting non renormalizable. In the UV limit, which corresponds to the infinite coupling limit, the 5d theory exhibits massless instanton operators which open up new directions of the Higgs branch of the moduli space of vacua.\\
In this paper, we expand the chiral ring at infinite coupling in terms of the U(1) instanton charge. Each sector is fully reproduced by the product of a bare instanton, defined by its flavour symmetry and R-charge representation, and a dressing factor common to all sectors. Such dressing factor is simply related to the chiral ring at finite coupling, with one additional colour in the gauge group.\\

\noindent We will consider $Sp(k)$ gauge theories, which have simpler Higgs branches than $SO$ and $SU$ theories, due to the absence of baryons. A further simplification with respect to unitary groups is the absence of a Chern-Simons level. There exists a discrete parameter due to $\pi_4(Sp(k))=\mathbb{Z}_2$, but it is irrelevant if massless flavours are present, as in our case.\\

\noindent We are going to encode the chiral ring in terms of the highest weight generating function (HWG), which enumerates operators transforming in the highest weight of the flavour symmetry, and grades them according to their R-charge. The characterization of the bare instanton, for any instanton charge, and the dressing factor was made possible by using the HWG function. Any other approach so far, including the Hilbert series, have not been successful in that regard.\\

\noindent For those combinations of colours and flavours that do not give symmetry enhancement in the UV, we will find the dressing factor as the residue of the Weyl integration of the HWG. In the few cases with symmetry enhancement, we will need to branch the UV representations in terms of the IR symmetries and subsequently perform the Weyl integration. We were able to analyse all combinations of $k$ and $N_f$, except $N_f=2k+5$, the case with the maximum amount of flavours allowing a 5d fixed point, corresponding to maximal enhancement of global symmetry $SO(2N_f)\to SO(2N_f+2)$.\\

\begin{figure}[H]
    \centering
  \begin{tikzpicture}[scale = 0.85,
  every node/.style={transform shape},
  box/.style={draw, rectangle, rounded corners, align=center, text width=5cm, minimum height=2.5cm},
  arrow/.style={->, thick},
  label/.style={midway, fill=white, inner sep=1pt, align=center},
  node distance=2.5cm and 5.5cm
]

\node[box] (A) at (0,0) {Chiral ring of Higgs branch\\ of 5d $\mathcal{N}=1$ Sp($k$)+$N_f$};
\node[box, right=of A] (B) {Moduli space of dressed\\ magnetic monopoles};
\node[box, below=of B] (C) {Moduli space of dressed\\ instanton operators};

\draw[arrow] (A) -- (B) 
  node[label, above] {Transverse movement of\\ 5-branes along 7-branes};

\draw[arrow] (B) -- (C) 
  node[label, right] {Residue computation in\\ instanton charge};

\draw[dotted, thick, ->] (C) to[out=150, in=-90] 
  node[label, below right=2pt and -12pt] {Moduli of\\ D1 inside D5} (A);

\end{tikzpicture}
    \caption{Logical flow of the paper}
    \label{fig:logicalflow}
\end{figure}
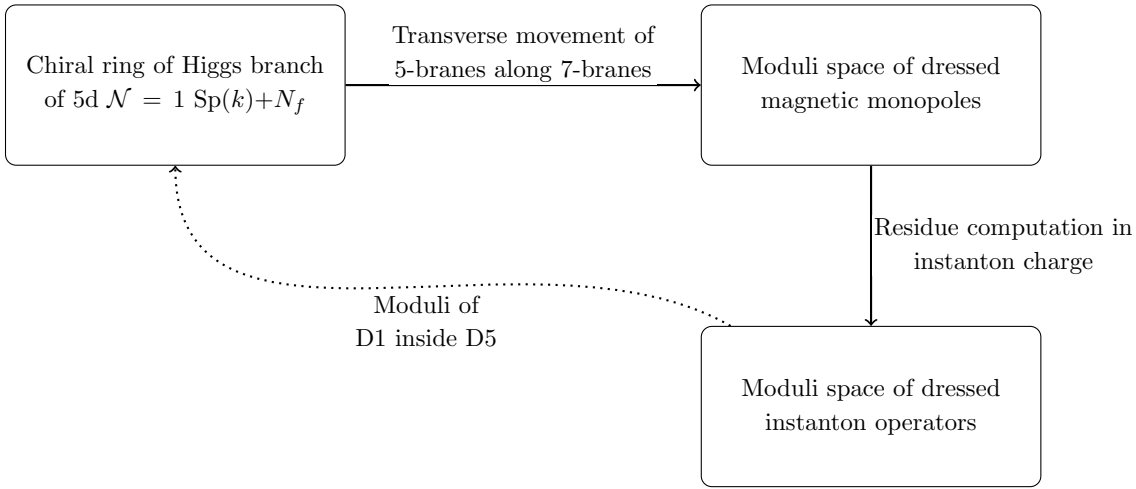

\noindent The logical flow that we will follow is represented in figure \ref{fig:logicalflow}.

\noindent The structure of the paper is the following: in section \ref{sec:field theory} we will review the moduli space of vacua of 5d $Sp(k)$ theories, and their chiral ring at classical and quantum level. In section \ref{sec:branedesciption} we will review the braneweb construction of the theories at hand, stressing the double interpretation of the Higgs branch as moduli space of dressed magnetic monopoles or dressed instantons. In section \ref{sec:classicalHiggs} we will comment on some results at classical level, while the original results of the paper are described in section \ref{sec:higgsatinfinitecoupling}, where we will explain the computations that have been performed and we will comment on their results.

\section{5d $\mathcal{N}=1$ $Sp(k)$ theories}\label{sec:field theory}
\noindent A peculiarity of 5d is the existence of a current built out of the field strength $F$ of the gauge vectors, which is topologically conserved due to the Bianchi identity. Namely:
\begin{equation}
    j= \star \:\text{tr}F \wedge F \ ,
\end{equation}
with the trace performed over colour indices. $j_{\mu}$ is the conserved current of a U(1) symmetry; fields charged under this U(1) are called instanton operators and as they increase or decrease a Hilbert state's instanton number by the instanton charge that they carry, hence they allow to move amongst topological sectors.\\

\noindent Instanton operators are, for what concerns this paper, 1/2 BPS local disorder operators, in the sense that they induce an instanton charge by imposing singularities in the path integral of the vector boson. As described in the seminal paper \cite{seiberg1996five}, the presence of massless instantons in the UV regime of the theory provides the existence of non trivial fixed points of the RG flow. These non trivial fixed points may enjoy an enhancement of the global symmetry (either the flavour symmetry or the spacetime isometry).\\

\noindent When the theory is superconformal, the current $j_{\mu}$ belongs to a short superconformal multiplet, where we also count the gaugino bilinear, which is a Lorentz scalar transforming in the $SU(2)_R$ spin 1 representation, with mass dimension $3$. The gaugino bilinear, or glueball, is a gauge-invariant holomorphic bilinear of the gaugino superfield. It is a chiral operator and it will play a crucial role in the coming discussion.\\

\noindent The quiver diagram of the theories we are considering is:

\begin{figure}[H]
    \centering
    \begin{tikzpicture}[
  roundnode/.style={circle, draw, minimum size=6pt, inner sep=0pt},
  squarenode/.style={rectangle, draw, minimum size=8pt, inner sep=2pt},
  label distance=1mm
]

  \node[roundnode] (Spk) at (0,0) {};
  \node[draw=none, below=1mm of Spk] {$Sp(k)$};

  \node[squarenode] (SO) at (2,0) {};
  \node[draw=none, below=1mm of SO] {$SO(2N_f)$};

  \draw[-] (Spk) -- (SO);

\end{tikzpicture} 
    \caption{Quiver diagram of $Sp(k)$ with $N_f$ flavours}
    \label{fig:quiverSpk}
\end{figure}

\noindent namely $Sp(k)$ gauge theory with $N_F$ flavours (or more precisely $2N_f$ half hypermultiplets, since the fundamental representation of $Sp(k)$ is pseudoreal). The round node represents a vector multiplet in the adjoint of $Sp(k)$, while the edge represents a hypermultiplet in the fundamental of $Sp(k)$ and vector of $SO(2N_f)$.\\

\noindent $Sp(k)$ theories have no Chern-Simons level. They have however a discrete level associated to $\pi_4(Sp(k))=\mathbb{Z}_2$, which is however irrelevant \cite{morrison1997extremal,bergman2014discrete} if massless flavours are present in the fundamental representation of $Sp(k)$. \noindent Unlike $SU(k)$ and $SO(k)$ gauge theories, $Sp(k)$ have no baryons, for any combination of flavours and colours. Indeed, while in the $SU$ or $SO$ cases there exists the group-invariant $\epsilon_{a_1\dots a_n}$ to contract and fully antisymmetrize the quarks' colour indices, in the $Sp$ case such invariant does not exist. This is the main reason why the Higgs branch of $Sp(k)$ theories is easier to study, with respect to the other gauge theories.

\subsection{Higgs branch of the moduli space of vacua}
\noindent The theory we are studying has two branches of moduli space of vacua, with possible non trivial mixed ones. There is a Coulomb branch, swept out by the vevs of the scalar field in the vector multiplet, and a Higgs branch, accounting for the vevs of the scalars in the hypermultiplets. In this paper we will only study the Higgs branch, both at finite and infinite coupling.\\

\noindent The classical Higgs branch is spanned by the gauge invariant vevs of scalars in the hypermultiplets obeying F and D constraint. We can consider gauge invariant scalar bilinears of the flavours:
\begin{equation}
    M_{[ij]}=Q_i^aQ_j^b\Omega_{ab}
\end{equation}
with $\Omega_{ab}$ being the $Sp(k)$ invariant, with $a$ colour indices $a=1\dots 2k$ and $i,j=1\dots 2N_f$ flavour indices. $Q_a^i$ is the scalar component of the hypermultiplet. Mesons transform in the spin 1 representation of the $SU(2)$ R-symmetry. They have $N_f(2N_f-1)$ complex components, but not all of them span the Higgs branch independently. Indeed, the classical Higgs branch only has complex dimensions as reported in table \ref{tab:cldimensions}:

\begin{table}[H]
    \centering
    \begin{tabular}{|c|c|} \hline
       Range &  Dim$_{\mathbb{C}}$   \\\hline
       $N_f \geq 2k+2$ & $2k(2N_f-2k-1)$ \\ \hline
         $N_f\leq 2k+1$  & $N_f(N_f-1)$ \\ \hline
    \end{tabular}
    \caption{Complex dimension of Higgs branch of $Sp(k)$ with $N_f$ flavours at finite coupling.}
    \label{tab:cldimensions}
\end{table}

\noindent Thus the mesons satisfy the following constraints:

\begin{equation}\label{eq:Josephrelations}
    \begin{array}{l}
        M^2=0 \\
        \text{rank}(M)\leq 2k \\
    \end{array}
\end{equation} 
where the first constraint implies that the rank of the meson matrix is at most $N_f$. This condition may result stronger or weaker than the second equation, according to the relative value of colours and flavours.\\ 
In particular \cite{Hanany:2016gbz}, $\mathcal{M}_{\text{Higgs}}^{\text{cl}}$ can be classified in all cases as a particular type of \textit{symplectic singularity}, known as \textit{closure of a nilpotent orbit} of $D_{N_f}$, denoted as $\bar{\mathcal{O}}_{D_{N_f}}^{[2^{2k},1^{2N_f-4k}]}$. \\

\noindent The mesons are the sole generators of the Higgs branch chiral ring in the IR. If flavours are absent, we do not have a continuous Higgs branch at finite coupling. Indeed the only other gauge invariant complex operator we can form is the gaugino superfield bilinear $S$, also known as glueball, which is classically nilpotent \cite{cremonesi2017instanton}:
\begin{equation}\label{eq:clrelationglueball}
    S^{h^{\vee}}=0 \ ,
\end{equation}
with $h^{\vee}$ the dual Coxeter number of the gauge group, which is $k+1$ for $Sp(k)$. Relation \ref{eq:clrelationglueball} is conjectured to imply that, in the absence of flavours, we have $h^{\vee}$ discrete vacua with a mass gap. \\

\noindent In the UV, the instantons become massless and add to the generators of the chiral ring and modify the existing constraints. Among these constraints, the gaugino bilinear nilpotency \ref{eq:clrelationglueball} is modified to:
\[
S^{h^{\vee}}= I \tilde{I} \ ,
\]
with $I$ and $\tilde{I}$ the instanton and the anti-instanton (i.e. of respective U(1) charge $1$ and -1). Such relation implies that the gaugino bilinear is massless too and necessarily belongs to the set of generators. In particular, it also implies that instantons of U(1) charge $1$ have R-charge $h^{\vee}/2$, since the gaugino bilinear has R-charge 1. There are no further generators besides the mesons, the instantons and the glueball.

\subsection{Hilbert series and Highest weight generating functions}
In order to enumerate the elements of the chiral ring we use \cite{benvenuti2007counting} the Hilbert series (HS), which grades the operators according to their R-charge:
\begin{equation}
    \text{HS}(\{x_i\},t)=\sum_{q=0}^{\infty}\chi(x_i)\cdot t^q \ .
\end{equation}
The meaning of this geometric series is that if an operator of the chiral ring transforms in the spin($n$) representation of the R-symmetry $SU(2)$, then it appears in the series at $t^{2n}$. The character of the representation under the global symmetry group is encoded in $\chi(x_i)$.\\

\noindent An alternative tool is that of the highest weight generating function (HWG) \cite{hanany2014highest}, which is analogous to the Hilbert series, except it replaces characters by highest weights of representations of the flavour symmetry. Therefore, the HWG offers an alternative when HS computations are unfeasible. 

\section{Brane description}\label{sec:branedesciption}
As pioneered in \cite{aharony1997branes, aharony1998webs}, 5d $\mathcal{N}=1$ theories can be studied as the worldvolume theories on webs of D5 branes 
 stretched between NS5 branes in Type IIB string theory.
The paradigmatic example is $SU(2)$:

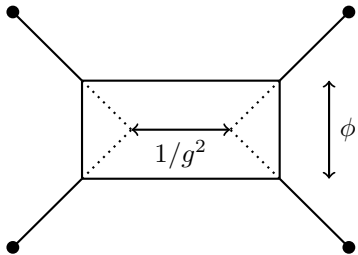
\begin{center}
\begin{tikzpicture}[scale=1.3, thick, every node/.style={font=\small},
    round/.style={circle, draw, fill=black, inner sep=0pt, minimum size=4pt}]

  \coordinate (A) at (0,0);
  \coordinate (B) at (2,0);
  \coordinate (C) at (2,1);
  \coordinate (D) at (0,1);
  \coordinate (E) at (0.5,0.5);
   \coordinate (F) at (1.5,0.5);
   \coordinate (G) at (2.5,0);
   \coordinate (H) at (2.5,1);

  \draw (A) -- (B);
  \draw (B) -- (C);
  \draw (C) -- (D);
  \draw (D) -- (A);

  \path (A) -- ++(-0.7,-0.7) coordinate (Aout);
  \path (B) -- ++(0.7,-0.7) coordinate (Bout);
  \path (C) -- ++(0.7,0.7) coordinate (Cout);
  \path (D) -- ++(-0.7,0.7) coordinate (Dout);

  \draw (A) -- (Aout) node[round] {};
  \draw (B) -- (Bout) node[round] {};
  \draw (C) -- (Cout) node[round] {};
  \draw (D) -- (Dout) node[round] {};

  \draw[dotted] (A) -- (E);
  \draw[dotted] (B) -- (F);
  \draw[dotted] (C) -- (F);
  \draw[dotted] (D) -- (E);

  \draw[<->, thick] (E) -- node[below]{$1/g^2$} (F) ;

  \draw[<->, thick] (H) -- node[right]{\(\phi\)} (G);

\end{tikzpicture}
\end{center}
\noindent with the vertical lines representing NS5 branes, the horizontal are D5 branes and the oblique lines are bound states of $(p,q)$ 5-branes. The black dots represent $[P,Q]$ 7-branes, and there is a residual $SL(2,\mathbb{Z})$ action on the coordinates of the plane. The vertical separation leads to adjoint Higgs mechanism for the worldvolume theory on the D5 branes, with $\phi$ the non vanishing vev of the Higgs field. When the D5 branes coincide, the horizontal distance corresponds to the squared inverse of the coupling constant, or equivalently to the mass of the gauge instantons. For many more details on branewebs we refer to the original papers.\\

\noindent The braneweb approach has been highly successful in studying the Higgs branch of vast classes of theories in the IR and in the UV (see for example \cite{bourget2020brane, cabrera2019tropical}). The main advantage of branewebs, in our context, is that despite the absence of a Lagrangian description in the UV, we can still take the limit of $1/g^2\to 0$, which in the braneweb corresponds to the horizontal separation of the NS5 branes. In that limit, we obtain a theory which does not depend on any lengthscale and which we thus assume to be superconformal\footnote{Of course scale invariance is proven to imply conformal invariance only in 2d. To prove invariance under the full superconformal group we would need a Lagrangian description, which is absent in the infinite coupling limit.}.
The braneweb describing $Sp(k)$ gauge theories with $N_f$ flavours is quite more elaborate than that for $SU(2)$. To our knowledge it was presented in \cite{bergman20155d}:

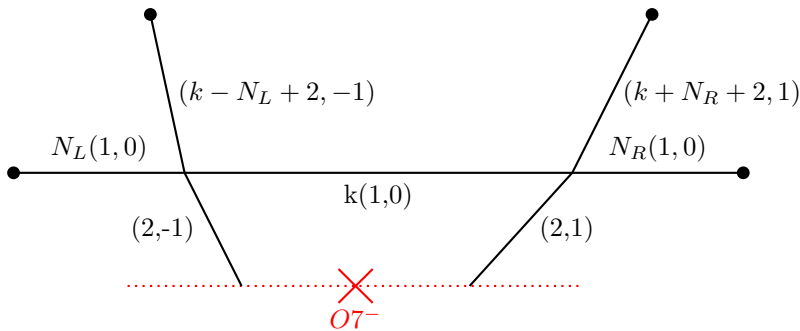
\begin{figure}[H]
    \centering
\begin{tikzpicture}[scale=1.5, thick, every node/.style={font=\small},
    round/.style={circle, draw, fill=black, inner sep=0pt, minimum size=4pt}]

  \coordinate (O7) at (0,0);
  \coordinate (L) at (-2,0);
  \coordinate (R) at (2,0);
  \coordinate (P1) at (-1,0);
  \coordinate (P2) at (1,0);
  \coordinate (N1) at (-1.5,1);
  \coordinate (N2) at (1.9,1);
  \coordinate (N22) at (2.6,2.4);
  \coordinate (N11) at (-1.8,2.4);
  \coordinate (N111) at (-3,1);
  \coordinate (N222) at (3.4,1);
  
  \draw[red, dotted, thick] (L) -- (R);

  \draw[red, thick] (-0.15,0.15) -- (0.15,-0.15);
  \draw[red, thick] (-0.15,-0.15) -- (0.15,0.15);
  \node[below=4pt] at (O7) {\textcolor{red}{$O7^-$}};

  \draw (P1) --  node[draw=none,left=1mm] {(2,-1)} (N1);
  
  \draw (P2) --  node[draw=none,right=1mm] {(2,1)} (N2) ;

  \draw[thick] (N1) -- node[draw=none,below] {k(1,0)} (N2);

  \draw (N1) -- node[draw=none,right] {$(k-N_L+2,-1)$} (N11) node[round] {};
  \draw (N2) -- node[draw=none,right] {$(k+N_R+2,1)$} (N22) node[round] {};

  \draw (N111) node[round] {} -- node[draw=none, above] {$N_L(1,0)$} (N1);
  \draw (N2) -- node[draw=none,above] {$N_R(1,0)$} (N222) node[round] {};
\end{tikzpicture}
\caption{Braneweb of Sp(k) gauge theory in 5d $\mathcal{N}=1$, with $N_f=N_L+N_R$ flavours}
\label{fig:braneweb}
\end{figure}

\noindent In Figure \ref{fig:braneweb}, the red dotted line represents the monodromy of the orientifold plane $O7^-$, depicted itself as a cross. Furthermore, the $N_f$ flavours are splitted arbitrarily into $N_L+N_R=N_f$.
The gauge group in the 5d theory on the D5 branes is $Sp(k)$, due to the orientifold plane, while the global symmetry group is $D_{N_F}$. The $O7^-$ plane reflects axially the braneweb below the monodromy line. The spacetime extension of the branes is:
\begin{table}[H]
    \centering
    \begin{tabular}{|c||l|} \hline
      Brane   & $ x^0 \qquad x^1  \qquad x^2  \qquad x^3  \qquad x^4 \qquad x^5 \qquad x^6 \qquad x^7 \qquad x^8  \qquad x^9$ \\ \hline \hline
       NS5  & $\times \qquad \times \qquad \times \qquad \times \qquad \times \qquad \times  \qquad \qquad  \qquad  \qquad $ \\ \hline
       D5 & $\times \qquad \times \qquad \times \qquad \times \qquad \times \qquad \qquad \quad \times \qquad  \qquad \qquad $ \\ \hline
       (p,q) 5-brane & $\times \qquad \times \qquad \times \qquad \times \qquad \times \qquad \; \; \text{angle }\alpha \qquad \qquad \qquad  \qquad  \qquad $ \\ \hline
       $O7^-$ & $\times \qquad \times \qquad \times \qquad \times \qquad \times \qquad  \qquad \qquad \qquad \times \qquad \times \qquad \times$ \\ \hline
       [P,Q] 7-brane & $\times \qquad \times \qquad \times \qquad \times \qquad \times \qquad  \qquad \qquad \qquad \times \qquad \times \qquad \times$ \\ \hline
    \end{tabular}
    \caption{Spacetime extension of braneweb.}
    \label{tab:my_label}
\end{table}

\noindent This configuration preserves 8 supercharges, i.e. $\mathcal{N}=1$. The spacetime isometry is broken to:

\begin{equation}
    SO(1,9)\to SO(1,4)\times SO(3)
\end{equation}
with $SO(1,4)$ being the isometry in the $x^{0\dots 4}$ directions, $SO(3)\sim SU(2)_R$ is the R-symmetry, i.e. rotations along $x^{7,8,9}$. The 5d $\mathcal{N}=1$ supermultiplets are classified in terms of the little group $SO(3)_L$ and the R-symmetry:

\begin{table}[h!]
    \centering
    \begin{tabular}{|c||c|c|} \hline 
        Supermultiplet & $SO(3)_L\times SU(2)_R$  & Content\\ \hline \hline
        Half-Hypermultiplet & [0;1]+[1;0] & 2 complex scalars and one fermion\\ \hline
        Vector multiplet & [1;1]+[2;0]+[0;0] & 1 fermion, gauge vector and real scalar\\ \hline
    \end{tabular}
    \caption{Simplest multiplets on the D5 worldvolume theory}
    \label{tab:5d multiplets}
\end{table}

\subsection{Computation of the Higgs branch chiral ring}
\noindent By the techniques explained in detail in \cite{cabrera2019tropical}, we are able to describe the Higgs branch of the theory as the moduli space of dressed magnetic monopole operators. Indeed, by considering consistent sub-webs and moving them along the 7-branes, we find that the low-energy degrees of freedom along the Higgs branch consist of virtual D3 branes stretched between the subwebs. As all D$(p-2)$ branes ending on D$p$ branes, the D3 branes act as 't Hooft Polyakov magnetic monopoles. Their moduli space is encoded in a quiver diagram, called a magnetic quiver, which can be evaluated through the monopole formula \cite{cremonesi2014monopole} to find the Higgs branch chiral ring.\\

\noindent The aim of the paper is to adopt a complementary point of view, where we express the Higgs branch of the D5 worldvolume as the moduli space of dressed instantons operators. A connection to the braneweb, where D1 inside D5 behave as gauge instantons, is however not yet established.\\

\noindent Besides the strings stretching between D1 and the 5-branes, there are also strings stretching between the D1 and the D7 branes, which are responsible for the gauge instanton zero modes. By quantization of the worldsheet fermions of the strings with mixed DN and ND boundary conditions it is possible to find the Clifford algebra representation in which the fermionic zero modes transforms. These zero modes do not change the instantons' energy, but increase their degeneracy and determine the $D_{N_f}$ representation to be the spinor representation in which they transform.

\section{Classical Higgs branch}\label{sec:classicalHiggs}
As we mentioned previously, at finite coupling the Higgs branch is parameterized only by the mesons. Its chiral ring is encoded in the highest weight generating functions reported in Table \ref{tab:finite coupling HWG and NOL}, where we summarise some results already contained in \cite{ferlito2016tale, bourget2020magnetic}. For completeness, we present in Table \ref{tab:magnetic quivers classical theory} the Higgs branches as moduli spaces of dressed magnetic monopoles, from which these HWG were computed.

\begin{table}[H]
    \centering
    \begin{tabular}{|c|c|c|} \hline
       Range  &  HWG & $\bar{\mathcal{O}}_D$ \\ \hline 
       & & \\
       $N_f \geq 2k+2$  & $\text{PE}\bigg[\sum_{i=1}^{k}\mu_{2i}t^{2i}\bigg]$ & $\bar{\mathcal{O}}_D^{[2^{2k},1^{2N_f-4k}]}$  \\
       &  & \\ \hline
       & & \\
      \makecell{$N_f \leq 2k+1$\\ $N_f$ even} & \makecell{$\text{PE}\bigg[\sum_{i=1}^{N_f/2-1}\mu_{2i}t^{2i}+(\mu^2_{N_f}+\mu^2_{N_f-1})t^{N_f}-$\\$-\mu^2_{N_f}\mu^2_{N_f-1}t^{2N_f}\bigg]$} & $\bar{\mathcal{O}}_D^{[2^{N_f}],I}+\bar{\mathcal{O}}_D^{[2^{N_f}],II}-\bar{O}_D^{[2^{N_f-2},1^4]}$ \\
      & & \\ \hline
      & & \\ 
     \makecell{ $N_f \leq 2k+1$\\ $N_f$ odd} & $\text{PE}\bigg[\sum_{i=1}^{(N_f-3)/2}\mu_{2i}t^{2i}+\mu_{N_f}\mu_{N_f-1}t^{N_f-1}\bigg]$& $\bar{\mathcal{O}}_D^{[2^{N_f-1},1^2]}$ \\
     & & \\ \hline
    \end{tabular}
    \caption{Highest weight generating functions of finite coupling Higgs branches of Sp(k) with $N_f$ flavours in the fundamental representation of the gauge group. The fugacities $\{\mu_i\}_{i=1\dots N_f}$ correspond to the flavour symmetry $D_{N_f}$. The third column classifies the singular varieties as a specific kind of symplectic singularity, namely closures of $D_{N_f}$ nilpotent orbits of height two. The two $\bar{\mathcal{O}}_D^{[2^{N_f}]}$, in the case $N_f\leq 2k+1$, $N_f$ even, represent the two cones that make up the Higgs branch.}
    \label{tab:finite coupling HWG and NOL}
\end{table}

\begin{table}[H]
    \centering
    \begin{tabular}{|m{2cm}|m{5cm}|m{3cm}|} \hline
      Range   & Quiver diagram & Notes \\ \hline
      & &  \\
      
         $N_f \geq 2k+2$ \vspace{0.4cm}
         & 
   \begin{tikzpicture}[
  scale=0.8, every node/.style={draw, circle, minimum size=4pt, inner sep=0pt},
  label distance=2mm
]

  \node (A1) at (0,0) {};
  \node (A2) at (1,0) {};
  \node (A3) at (2,0) {};
  \node (A4) at (3,0) {};
  \node (A5) at (4,0) {};

  \node[draw=none, below=1mm of A1] {1};
  \node[draw=none, below=1mm of A2] {2};
  \node[draw=none, below=-2mm of A3] {$2k{-}1$};
  \node[draw=none, below=1mm of A4] {$2k$};
  \node[draw=none, below=1mm of A5] {$2k$};

  \node[draw=none] at (1.5,0) {$\cdots$};
  \node[draw=none] at (3.5,0) {$\cdots$};

  \draw (A1) -- (A2);
  \draw (A3) -- (A4);

  \node (Top) at (3,1) {};
  \node[draw=none, above=1mm of Top] {1};
  \draw (A4) -- (Top);

  \node (NE) at (5,0.7) {};
  \node (SE) at (5,-0.7) {};
  \node[draw=none, right=1mm of NE] {k};
  \node[draw=none, right=1mm of SE] {k};
  \draw (A5) -- (NE);
  \draw (A5) -- (SE);

   \coordinate (B1) at (2.9,-1);
  \coordinate (B2) at (4.1,-1);
  \draw [decorate, decoration={brace, amplitude=6pt, mirror}, thick]
    (B1) -- (B2)
    node[draw=none,midway, below=-9pt] {\( N_f - 2k - 1 \)};
\end{tikzpicture} & 
        \\  \hline
     \makecell{$N_f \leq 2k+1$\\ $N_f$ even} & \begin{tikzpicture}[
  every node/.style={draw, circle, minimum size=4pt, inner sep=0pt},
  label distance=1mm
]

  \node (A1) at (0,0) {};
  \node (A2) at (2,0) {}; 

  \node[draw=none, below=1mm of A1] {1};
  \node[draw=none, right=2mm of A2] {$N_f - 2$};

  \draw[dotted, thick] (A1) -- (A2);

  \node (NE) at (3,0.7) {};
  \node (SE) at (3,-0.7) {};
  \node[draw=none, above=-1mm of NE] {$N_f/2$};
  \node[draw=none, below=-3mm of SE] {$\frac{N_f}{2} - 1$};

  \draw (A2) -- (NE);
  \draw (A2) -- (SE);

  \node (Right) at (4.2,0.7) {};
  \node[draw=none, right=1mm of Right] {1};
  \draw (NE) -- node[draw=none,below=1mm] {2} (Right);

\end{tikzpicture} & Cone I\\
& \begin{tikzpicture}[
  every node/.style={draw, circle, minimum size=4pt, inner sep=0pt},
  label distance=1mm
]

  \node (A1) at (0,0) {};
  \node (A2) at (2,0) {}; 

  \node[draw=none, below=1mm of A1] {1};
  \node[draw=none, right=2mm of A2] {$N_f - 2$};

  \draw[dotted, thick] (A1) -- (A2);

  \node (NE) at (3,0.7) {};
  \node (SE) at (3,-0.7) {};
  \node[draw=none, above=-5mm of NE] {$N_f/2-1$};
  \node[draw=none, below=-3mm of SE] {$N_F/2$};

  \draw (A2) -- (NE);
  \draw (A2) -- (SE);

  \node (Right) at (4.2,-0.7) {};
  \node[draw=none, right=1mm of Right] {1};
  \draw (SE) -- node[draw=none,above=1mm] {2} (Right);

\end{tikzpicture} & Cone II\\
& \begin{tikzpicture}[
  every node/.style={draw, circle, minimum size=4pt, inner sep=0pt},
  label distance=1mm
]

  \node (A1) at (0,0) {};
  \node (A2) at (2,0) {}; 

  \node[draw=none, below=1mm of A1] {1};
  \node[draw=none, right=2mm of A2] {$N_f - 2$};

  \draw[dotted, thick] (A1) -- (A2);

  \node (NE) at (3,0.7) {};
  \node (SE) at (3,-0.7) {};
  \node[draw=none, above=-4mm of NE] {$N_f/2-1$};
  \node[draw=none, below=-3mm of SE] {$\frac{N_f}{2} - 1$};

  \draw (A2) -- (NE);
  \draw (A2) -- (SE);

  \node (Right) at (2,0.7) {};
  \node[draw=none, left=1mm of Right] {1};
  \draw (A2) -- node[draw=none] {} (Right);

\end{tikzpicture} & Intersection\\ \hline 
      \makecell{ $N_f \leq 2k+1$\\ $N_f$ odd} & 
      \begin{tikzpicture}[
  every node/.style={draw, circle, minimum size=4pt, inner sep=0pt},
  label distance=1mm
]

  \node (A1) at (0,0) {};
  \node (A2) at (2,0) {}; 

  \node[draw=none, below=1mm of A1] {1};
  \node[draw=none, right=2mm of A2] {$N_f - 2$};

  \draw[dotted, thick] (A1) -- (A2);

  \node (NE) at (3.3,1.2) {};
  \node (SE) at (3.3,-1.2) {};
  \node[draw=none, above=-5mm of NE] {$(N_f-1)/2$};
  \node[draw=none, below=-3mm of SE] {$(N_F-1)/2$};

  \draw (A2) -- (NE);
  \draw (A2) -- (SE);

  \node (Right) at (4.5,0) {};
  \node[draw=none, right=1mm of Right] {1};
  \draw (SE) -- node[draw=none] {} (Right);
  \draw (NE) -- node[draw=none] {} (Right);
\end{tikzpicture}
      & \\ \hline
    \end{tabular}
    \caption{Quiver diagrams encoding the Higgs branches at finite coupling of Sp(k) with $N_f$ flavours as moduli spaces of dressed magnetic monopoles (i.e. \textit{magnetic quivers}). The dotted line represents gauge groups increasing linearly from one extremum to the other. For even $N_f$ and $N_f < 2k+1 $ the Higgs branch at finite coupling is the union of two cones.}
    \label{tab:magnetic quivers classical theory}
\end{table}
   
\noindent For any non vanishing number of flavours, part of the the Higgs branch chiral ring is generated, at the level of highest weights, by even rank tensors of $D_{N_f}$, with bounded rank. These correspond to the mesons in the representation $\mu_2$ and representation in its symmetric product $\mu_4,\mu_6\dots$. In particular, for high number of flavours ($N_f\geq 2k+2$) this is the only contribution to the Higgs branch chiral ring. The bound on the rank of mesons representation comes from the number of flavours, namely\footnote{In eq. \ref{eq:bound on the meson rank}, $\lfloor \:\: \rfloor$ denotes the floor function, which rounds the number to the closest inferior integer.}:
\begin{equation}\label{eq:bound on the meson rank}
    \mu_{2i} \quad \text{s.t.} \quad i\leq \lfloor N_f/2\rfloor-1 \ .
\end{equation}
\\
Furthermore, for $N_f \leq 2k+1$ and $N_f$ even, the Higgs branch is composed of two isomorphic cones with a nonvanishing intersection, whose only distinction is due to the $D_{N_f}$ spinor type. One cone has HWG function:

\begin{equation}\label{eq:hwg single cone}
    \text{HWG}_{\text{single cone}}=\text{PE}\bigg[\sum_{i=1}^{N_f/2-1}\mu_{2i}t^{2i}+\mu_{N_f}^{2}t^{N_f}\bigg]
\end{equation}
and the other has the same HWG, with the other spinor. The intersection is itself the closure of a nilpotent orbit, with HWG:
\begin{equation}
    \text{HWG}_{\text{intersection}}=\text{PE}\bigg[\sum^{N_f/2-1}_{i=1}\mu_{2i}t^{2i}\bigg]
\end{equation}
Moreover, for odd $N_f$, the HWG is the sum of the meson contribution and the $(N_f-1)$-th antisymmetric product of the fundamental of $D_{N_f}$:
\begin{equation} \label{odd flavour condensate at finite coupling}
    \wedge^{(N_f-1)}[1,0,\dots,0]=[0,\dots,0,1,1] \ ,
\end{equation}
or in terms of highest weights:
\begin{equation} 
    \wedge^{(N_f-1)}\mu_1=\mu_{N_f-1}\mu_{N_f} \ ,
\end{equation}
Similarly, for even flavours:
\begin{equation}\label{even flavour condensate at finite coupling}
    \wedge^{(N_f)}[1,0,\dots,0]=[0,\dots,0,2]+[0,\dots0,2,0] \ ,
\end{equation}
and in terms of highest weights:
\begin{equation}
    \wedge^{(N_f)}\mu_1=\mu_{N_f}^2+\mu_{N_f-1}^2 \ .
\end{equation}
Both representations \ref{even flavour condensate at finite coupling} and \ref{odd flavour condensate at finite coupling} appear in their HWG at R-charge $2k$. As we will see shortly (in section \ref{par:bareinstanton}), an alternative description of \ref{odd flavour condensate at finite coupling} and \ref{even flavour condensate at finite coupling} is that of bound states of one instanton and one anti-instanton.

\section{Higgs branch at infinite coupling}\label{sec:higgsatinfinitecoupling}
The theories we are considering, 5d $\mathcal{N}=1$ $Sp(k)$ gauge theories, do not have a Lagrangian formulation at infinite coupling, making the study of its moduli space of vacua non trivial. One way to perform such study is by adopting the tool of branewebs, discussed in a previous paragraph. Branewebs provide the combinatorial data encoding the moduli space of vacua of the theory. In particular, the infinite coupling limit has a clear interpretation in terms of branewebs, corresponding to the limit of vanishing horizontal distance of the NS5 branes (in the case of $Sp(k)$, they are more properly $(p,1)$ 5 branes). It is possible to deform Figure \ref{fig:braneweb} to infinite coupling as long as $N_f\leq 2k+5$, otherwise external legs intersect, leading to a configuration with additional massless degrees of freedom (namely the strings stretching between the finite branes) representing a different gauge theory altogether \cite{kim2015tao}.\\
As mentioned earlier, if $N_f=2k+6$ flavours are present, the theory has a 6d fixed point, i.e. there is an enhancement of the Lorentz symmetry instead of the flavour symmetry. For $N_f>2k+6$ the UV regime of the theory is still scarcely understood.

\subsection{Highest weight generating functions at infinite coupling}

As we described earlier, at infinite coupling the Higgs branch is parameterized by mesons, instantons, anti-instantons and gaugino bilinears. The growth in dimension of the Higgs branch is reported in table \ref{tab:infdimensions}.
The HWG functions of the Higgs branch chiral ring are found to correspond to the exceptional sequences detailed in \cite{cremonesi2017instanton} (except when $N_f \leq 2k+1$), and are reported in Table \ref{tab:hwg at infinite coupling}. As mentioned previously, the only difference between the $N_f$ even and odd case for $N_f<2k+1$ concerns the properties of spinors for $SO(2N_f)$ for even and odd rank, but they have a unified description as the $2k$-th antisymmetric product of the fundamental representation. The corresponding quivers are reported in Table \ref{tab:magneticquiversinfinitecoupling}. 

\begin{table}[H]
    \centering
    \begin{tabular}{|c|c|c|} \hline
       Flavours &  Dim$_{\mathbb{C}}$ at finite coupling &  Dim$_{\mathbb{C}}$ at infinite coupling  \\\hline
       $N_f = 2k+5$ & $2k(2N_f-2k-1)$ & $N_f(N_f+1)+1$ \\ \hline
         $N_f =2k+4$  & $2k(2N_f-2k-1)$ & $N_f(N_f-1)+2$ \\ \hline
         $2k+2 \leq N_f \leq 2k+3$ &$2k(2N_f-2k-1)$ & $N_f(N_f-1)+1$ \\ \hline
        $N_f\leq 2k+1$  & $N_f(N_f-1)$ &  $N_f(N_f-1)+1$ \\ \hline
    \end{tabular}
    \caption{Complex dimension of Higgs branch of $Sp(k)$ with $N_f$ flavours at finite and infinite coupling.}
    \label{tab:infdimensions}
\end{table}

\begin{table}[H]
    \centering
    \begin{tabular}{|c|c|c|} \hline
       Range  & Flavour group & HWG \\ \hline
       & & \\
        $N_f=2k+5$ & $SO(2N_f+2)$ & $\text{PE}\bigg[\sum_{i=1}^{k+2}\mu_{2i}t^{2i}+t^4+\mu_{2k+6}(t^{k+1}+t^{k+3})\bigg]$ \\
        & & \\ \hline
        & & \\ 
        $N_f=2k+4$ & $SO(2N_f)\times SU(2)$ & \makecell{$\text{PE}\bigg[\sum_{i=1}^{k+1}\mu_{2i}t^{2i}+t^4+\nu^2 t^2+\nu\mu_{2k+4}(t^{k+1}+t^{k+3})+$\\$+\mu_{2k+4}^2t^{2k+4}-\nu^2 \mu^2_{2k+4}t^{2k+6}\bigg]$} \\
        & & \\ \hline 
        & & \\
       $N_f =2k+3$ & $SO(2N_f)\times U(1)$ & $\text{PE}\bigg[\sum_{i=1}^{k}\mu_{2i}t^{2i}+t^2+(q\mu_{2k+2}+q^{-1}\mu_{2k+3})t^{k+1} \bigg]$ \\
        & & \\ \hline
        & & \\
        $N_f=2k+2$ & $SO(2N_f)\times U(1)$ & $\text{PE}\bigg[\sum_{i=1}^{k}\mu_{2i}t^{2i}+t^2+(q+q^{-1})\mu_{2k+2}t^{k+1}\bigg]$ \\ 
        & & \\ \hline
        &  & \\
       $N_f=2k+1$ & $SO(2N_f)\times U(1)$ & \makecell{$\text{PE}\bigg[\sum_{i=1}^{k-1}\mu_{2i}t^{2i}+\mu_{2k+1}\mu_{2k}t^{2k}+t^2+$\\$+t^{k+1}(q\mu_{2k}+q^{-1}\mu_{2k+1})-\mu_{2k+1}\mu_{2k}t^{2(k+1)}\bigg]$} \\
      & & \\ \hline
      & & \\
      \makecell{$N_f\leq 2k$\\ $N_f$ even} & $SO(2N_f)\times U(1)$ & \makecell{$\text{PE}\bigg[\sum_{i=1}^{N_f/2-1}\mu_{2i}t^{2i}+\mu_{N_f-1}^2t^{N_f}\bigg]+$\\$+\text{PE}\bigg[\sum_{i=1}^{N_f/2-1}\mu_{2i}t^{2i}+\mu_{N_f}^2t^{N_f}+$\\$+t^2+(q+q^{-1})\mu_{N_f}t^{k+1}-\mu_{N_f}^2t^{2(k+1)}\bigg]-$ \\ $-\text{PE}\bigg[\sum_{i=1}^{N_f/2-1}\mu_{2i}t^{2i}\bigg]$} \\ 
      & & \\ \hline
      & & \\
        \makecell{$N_f < 2k+1$\\ $N_f$ odd} & $SO(2N_f)\times U(1)$ & \makecell{$\text{PE}\bigg[\sum_{i=1}^{(N_f-1)/2-1}\mu_{2i}t^{2i}+\mu_{N_f}\mu_{N_f-1}t^{N_f-1}+t^2+$\\$+(q\mu_{N_f-1}+q^{-1}\mu_{N_f})t^{k+1}-\mu_{N_f}\mu_{N_f-1}t^{2(k+1)}\bigg]$}  \\
        & & \\ \hline
    \end{tabular}
    \caption{HWGs for chiral rings at infinite coupling of Higgs branch of Sp(k) with $N_f$ flavours in the fundamental of the gauge group. The fugacities $\mu_i$ are related to the D-type part of the flavour group, while $\nu$ is a SU(2) fugacity and $q$ is the U(1) charge.}
    \label{tab:hwg at infinite coupling}
\end{table}

\begin{longtable}{|>{\centering\arraybackslash}m{2.5cm}|l|>{\centering\arraybackslash}m{2.5cm}|} \hline
        Range & Quiver & Notes \\ \hline
        $N_f=2k+5$  &  
\begin{tikzpicture}[
  scale=1, every node/.style={draw, circle, minimum size=5pt, inner sep=0pt},
  label distance=2mm
]

  \node (A1) at (0,0) {};
  \node (A4) at (2,0) {};
  \node (A5) at (3.5,0) {};

  \node[draw=none, below=1mm of A1] {1};
  \node[draw=none, below=-3mm of A4] {$2k+4$};
  \node[draw=none, above=-1mm of A5] {$k+3$};
  
  \draw[dotted] (A1) -- (A4);

  \node (Top) at (2,1) {};
  \node[draw=none, above=-1mm of Top] {$k+2$};
  \draw (A4) -- (Top);

  \node (NE) at (5,0) {};
  \node[draw=none, above=1mm of NE] {2};
  \draw (A4) -- (A5);
  \draw (A5) -- (NE);
\end{tikzpicture} & $E_8$ family\\ \hline
     $N_f=2k+4$  & \begin{tikzpicture}[
  scale=1, every node/.style={draw, circle, minimum size=5pt, inner sep=0pt},
  label distance=2mm
]

  \node (A1) at (0,0) {};
  \node (A4) at (2,0) {};
  \node (A5) at (3.5,0) {};

  \node[draw=none, below=1mm of A1] {1};
  \node[draw=none, below=-3mm of A4] {$2k+2$};
  \node[draw=none, above=-1mm of A5] {$k+2$};
  
  \draw[dotted] (A1) -- (A4);

  \node (Top) at (2,1) {};
  \node[draw=none, above=-1mm of Top] {$k+1$};
  \draw (A4) -- (Top);

  \node (NE) at (5,0) {};
  \node[draw=none, above=1mm of NE] {2};
  \draw (A4) -- (A5);
  \draw (A5) -- (NE);

   \node (NU) at (6.5,0) {};
  \node[draw=none, above=1mm of NU] {1};
  \draw (NE) -- (NU);
\end{tikzpicture} & $E_7$ family\\  \hline \vspace{1cm}
$N_f=2k+3$ & 
 \begin{tikzpicture}[
  scale=1, every node/.style={draw, circle, minimum size=5pt, inner sep=0pt},
  label distance=2mm
]

  \node (A1) at (0,-1) {};
  \node (A4) at (2,-1) {};
  \node (A5) at (3.5,-1) {};
  \node (A6) at (2,1) {};

  \node[draw=none, below=1mm of A1] {1};
  \node[draw=none, below=-3mm of A4] {$2k+1$};
  \node[draw=none, above=-1mm of A5] {$k+1$};
  
  \draw[dotted] (A1) -- (A4);

  \node (Top) at (2,0) {};
  \node[draw=none, left=0mm of Top] {$k+1$};
  \draw (A4) -- (Top);
\node[draw=none, above=5mm of A6] {};
  \node[draw=none, above=1mm of A6] {$1$};
  \draw (Top) -- (A6);

  \node (NE) at (5,-1) {};
  \node[draw=none, below=1mm of NE] {1};
  \draw (A4) -- (A5);
  \draw (A5) -- (NE);
\end{tikzpicture}
& $E_6$ family\\ \hline
$N_f=2k+2$  & \begin{tikzpicture}[
  scale=1, every node/.style={draw, circle, minimum size=5pt, inner sep=0pt},
  label distance=2mm
]

  \node (A1) at (0,-1) {};
  \node (A4) at (2,-1) {};
  \node (A5) at (3.5,-1) {};
  \node (A6) at (3.5,0) {};

  \node[draw=none, below=1mm of A1] {1};
  \node[draw=none, below=0mm of A4] {$2k$};
  \node[draw=none, below=-2mm of A5] {$k+1$};
  
  \draw[dotted] (A1) -- (A4);

  \node (Top) at (2,0) {};
  \node[draw=none, above=0mm of Top] {$k$};
  \draw (A4) -- (Top);
\node[draw=none, above=5mm of A6] {};
  \node[draw=none, above=1mm of A6] {$1$};
  \draw (A5) -- (A6);

  \node (NE) at (5,-1) {};
  \node[draw=none, below=1mm of NE] {1};
  \draw (A4) -- (A5);
  \draw (A5) -- (NE);
\end{tikzpicture} & $E_5$ family \\ \hline
$N_f=2k+1$  &  
 \begin{tikzpicture}[
  scale=1, every node/.style={draw, circle, minimum size=5pt, inner sep=0pt},
  label distance=2mm
]

  \node (A1) at (0,-1) {};
  \node (A4) at (2,-1) {};
  \node (A5) at (3.5,-1) {};
  \node (A6) at (2,1) {};

  \node[draw=none, below=1mm of A1] {1};
  \node[draw=none, below=-3mm of A4] {$2k-1$};
  \node[draw=none, below=0mm of A5] {$k$};
  
  \draw[dotted] (A1) -- (A4);

  \node (Top) at (2,0) {};
  \node[draw=none, left=0mm of Top] {$k$};
  \draw (A4) -- (Top);
\node[draw=none, above=5mm of A6] {};
  \node[draw=none, left=1mm of A6] {$1$};
  \draw (Top) -- (A6);

  \node (NE) at (5,-1) {};
  \node[draw=none, below=1mm of NE] {1};
  \draw (A4) -- (A5);
  \draw (A5) -- (NE);
  \draw (A6) -- (NE);
\end{tikzpicture}
& $E_4$ family \\ \hline
\makecell{$N_f<2k+1$ \\ $N_f$ even}  &
 \begin{tikzpicture}[
  scale=1, every node/.style={draw, circle, minimum size=5pt, inner sep=0pt},
  label distance=2mm
]

  \node (A1) at (0,-1) {};
  \node (A4) at (2,-1) {};
  \node (A5) at (3.5,-1) {};
  \node (A6) at (2,1) {};

  \node[draw=none, below=1mm of A1] {1};
  \node[draw=none, below=-3mm of A4] {$N_f-2$};
  \node[draw=none, right=0mm of A5] {$N_f/2-1$};
  
  \draw[dotted] (A1) -- (A4);

  \node (Top) at (2,0) {};
  \node[draw=none, left=0mm of Top] {$N_f/2$};
  \draw (A4) -- (Top);
\node[draw=none, above=5mm of A6] {};
  \node[draw=none, left=1mm of A6] {$1$};
  \draw (Top) -- node[draw=none,right] {2} (A6);

 
  \draw (A4) -- (A5);

\end{tikzpicture}
& Cone I\\
&
\begin{tikzpicture}[
  scale=1, every node/.style={draw, circle, minimum size=5pt, inner sep=0pt},
  label distance=2mm
]

  \node (A1) at (0,-1) {};
  \node (A4) at (2,-1) {};
  \node (A5) at (3.5,-1) {};
  \node (A6) at (3.5,0) {};
  \node (A7) at (5,-1) {};

  \node[draw=none, below=0mm of A7] {1};

  \node[draw=none, below=1mm of A1] {1};
  \node[draw=none, below=-1mm of A4] {$N_f-2$};
  \node[draw=none, below=0mm of A5] {$N_f/2$};
  
  \draw[dotted] (A1) -- (A4);

  \node (Top) at (2,0) {};
  \node[draw=none, above=-5mm of Top] {$N_f/2-1$};
  \draw (A4) -- (Top);
\node[draw=none, above=5mm of A6] {};
  \node[draw=none, above=1mm of A6] {$1$};
  \draw (A5) -- (A6);

  \draw (A4) -- (A5) -- (A7);
  \draw (A6) -- (A7);

  \node[draw=none] (A8) at (4.2,-0.3) {};
  \node[draw=none, right=0mm of A8] {$k+1-N_f/2$};
\end{tikzpicture}
& Cone II \\
& 
 \begin{tikzpicture}[
  scale=1, every node/.style={draw, circle, minimum size=5pt, inner sep=0pt},
  label distance=2mm
]

  \node (A1) at (0,-1) {};
  \node (A4) at (2,-1) {};
  \node (A5) at (3.5,-1) {};
  \node (A6) at (3.5,0) {};

  \node[draw=none, below=1mm of A1] {1};
  \node[draw=none, below=-3mm of A4] {$N_f-2$};
  \node[draw=none, right=0mm of A5] {$N_f/2-1$};
  
  \draw[dotted] (A1) -- (A4);

  \node (Top) at (2,0) {};
  \node[draw=none, above=-5mm of Top] {$N_f/2-1$};
  \draw (A4) -- (Top);
\node[draw=none, above=5mm of A6] {};
  \node[draw=none, right=1mm of A6] {$1$};
  \draw (A4) -- node[draw=none,left] {} (A6);

  \draw (A4) -- (A5);

\end{tikzpicture}
& Intersection  \\ \hline
 \makecell{$N_f<2k+1$ \\ $N_f$ odd} \vspace{2cm} & 
 \begin{tikzpicture}[
  scale=1, every node/.style={draw, circle, minimum size=5pt, inner sep=0pt},
  label distance=2mm
]

  \node (A1) at (0,-1) {};
  \node (A4) at (2,-1) {};
  \node (A5) at (4,-1) {};
  \node (A6) at (2,1) {};
  \node[draw=none] (A7) at (3.2,0.4) {};

  \node[draw=none, right=1mm of A7] {$k+1-(N_f-1)/2$};

  \node[draw=none, below=1mm of A1] {1};
  \node[draw=none, below=-3mm of A4] {$N_f-2$};
  \node[draw=none, below=-7mm of A5] {$(N_f-1)/2$};
  
  \draw[dotted] (A1) -- (A4);

  \node (Top) at (2,0) {};
  \node[draw=none, left=0mm of Top] {$(N_f-1)/2$};
  \draw (A4) -- (Top);
\node[draw=none, above=5mm of A6] {};
  \node[draw=none, left=1mm of A6] {$1$};
  \draw (Top) -- (A6);

  \node (NE) at (6,-1) {};
  \node[draw=none, right=1mm of NE] {1};
  \draw (A4) -- (A5);
  \draw (A5) -- (NE);
  \draw (A6) -- (NE);
\end{tikzpicture}
& \vspace{-2cm}\\ \hline
\caption{Quiver diagrams encoding the Higgs branch at infinite coupling as moduli space of dressed monopole operators for $Sp(k)$ with $N_f$ flavours. The dotted line represents gauge groups increasing linearly from one extremum to the other.}  \label{tab:magneticquiversinfinitecoupling}\\
\end{longtable}

\noindent As Table \ref{tab:magneticquiversinfinitecoupling} reports, the theories we are considering exhibit global symmetry enhancement only for two combinations of flavour and colours: $N_f=\{2k+4,2k+5\}$. In all other cases, the global symmetry in the UV matches exactly the one in the IR, namely $SO(2N_f)\times U(1)$, with $U(1)$ being the instanton number symmetry.\\

\noindent For the case of $N_f<2k+1$ and $N_f$ even, the intersection can be found by repeated quiver subtraction \cite{cabrera2018quiver} from quivers of each cone. In particular, while at finite coupling it is sufficient to subtract an affine $a_1$ Dynkin diagram from each quiver in order to find the intersection, the subtraction is more elaborate in the limit of infinite coupling (unless $k=N_f/2$, as examined in \cite{ferlito2016tale, Bourget:2023cgs}). Indeed, we need to subtract from Cone II an affine $a_{k-N_f/2}$ times, followed by a single affine $a_1$. What this means, physically, is that at in the UV new phases may appear, characterized by new patterns of Higgs breaking of the gauge symmetry.\\

\noindent Quiver subtraction allows to study the stratification of the Higgs branch in symplectic singularities, encoded in the Hasse diagram \cite{bourget2020higgs}. Physically, such stratification reproduces the pattern of Higgs breaking of the gauge group in different loci of the moduli space, for theories with a Lagrangian description. The top part of the Hasse diagram for $N_f<2k+1$ with $N_f$ even, has the following finite-infinite coupling enhancement:

\begin{table}[H]
    \centering
    \begin{tabular}{|c|c|} \hline
        Finite coupling & Infinite coupling \\ \hline 
        & \\
    \begin{tikzpicture}[scale=1, every node/.style={scale=1}]

  \filldraw (0,2) circle (2pt);          
  \filldraw (2,2) circle (2pt);          

  \filldraw (1,1) circle (2pt);

  \filldraw (1,0) circle (2pt);

  \draw (0,2) -- (1,1) node[midway, left] {$a_1$};
  \draw (2,2) -- (1,1) node[midway, right] {$a_1$};
  \draw (1,1) -- (1,0) node[midway, right] {$d_{4}$};

  \draw (1,0) -- (1,-0.5);
  \node at (1,-0.8) {$\cdots$};

\end{tikzpicture} & 
 \begin{tikzpicture}[scale=1, every node/.style={scale=1}]

  \filldraw (0,2) circle (2pt);          
  \filldraw (2,2) circle (2pt);          

  \filldraw (3,3) circle (2pt);          

  \filldraw (1,1) circle (2pt);

  \filldraw (1,0) circle (2pt);

  \draw (0,2) -- (1,1) node[midway, left] {$a_1$};
  \draw (2,2) -- (1,1) node[midway, right] {$a_1$};
  \draw (1,1) -- (1,0) node[midway, right] {$d_{4}$};
  \draw (2,2) -- (3,3) node[midway, right] {$A_{k-N_f/2}$};

  \draw (1,0) -- (1,-0.5);
  \node at (1,-0.8) {$\cdots$};

\end{tikzpicture}\\
& \\ \hline
    \end{tabular}
    \caption{Hasse diagrams of Sp(k) with $N_f<2k+1$, $N_f$ even, at finite and infinite coupling.}
    \label{tab:my_label}
\end{table}
\noindent In the case $k=N_f/2$, the $a_1$ and the enhanced slice are substituted by a unique $a_2$. This case only shows up for $N_f$ even.\\

\noindent We will now divide the Higgs branch chiral ring into infinite topological sectors, corresponding to different U(1) charge, and write the HWG function for each sector. Afterwards, we will show that taking the sum over all highest weight generating functions of given sectors, we will recover the known HWG of Table \ref{tab:hwg at infinite coupling}.

\subsection{Dressed instantons}\label{par:dressedinstantons}
In the course of this section we will divide the Higgs branch chiral ring into infinite topological sectors. Each sector of charge $I$ will be denoted by a bare instanton, expressing the lower R-charge and smallest $D_{N_f}$ representation at which an instanton of charge $I$ arises. Each topological sector is swept out by the product of the bare instanton and a dressing factor, which adds R-charge and $D_{N_f}$ representations to the bare instanton. Namely, we will write the HWG of the Higgs branch chiral ring in the sector $I$, denoted as $\text{HWG}_I$, as the product:
\[
\text{HWG}_I=(\text{bare instanton of charge $I$)}\cdot \text{(dressing)} \ .
\]
As an example, we will first consider the case $N_f=2k+1$ and show that its Higgs branch (at the level of highest $D_{N_f}$ weights) can be written as the moduli space of dressed instantons. It is easier to first consider a case without symmetry enhancement, as the U(1) instanton number is unaltered in the UV.\\

\paragraph{Dressing factor}
We can find the dressing factor by computing the residue around poles in $q$ of the HWG function, namely by performing a Hyperkaehler quotient with respect to U(1), as follows:
\begin{equation}
    \text{HWG}_0=\int \dfrac{\text{d}q}{q}\cdot \text{HWG}\cdot \text{PE}\bigg[-t^2\bigg] \ .
\end{equation}
Let us explain this formula:
\begin{itemize}
\item The measure $\text{d}q/q$ is the Haar measure for the group U(1),
\item The Weyl integration of the flavour group U(1) selects the charge-independent part of the HWG, as it corresponds to the D-term constraint,
\item The $\text{PE}[-t^2]$ corresponds to the F-term constraint, and has the effect of removing from the HWG function the gaugino bilinear, which transforms in the same supermultiplet as the instanton current, as mentioned previously. Therefore, in order to preserve supersymmetry the F-term constraint and D-term constraint have to be performed together. The coefficient of $t^2$ correspond to the adjoint representation of the instanton symmetry; in our case the instanton symmetry is U(1), hence the adjoint representation is trivial. However, there will be one case with enhanced SU(2) instanton symmetry, where the Hyperkaehler quotient is performed with the adjoint of $SU(2)$, namely $[2]$.
\end{itemize}
The dressing factor can be computed to be:
\begin{equation}\label{eq:2k+1 weyl integration}
    \text{HWG}_0=\text{PE}\bigg[\sum_{i=1}^{k-1}\mu_{2i}t^{2i}+\mu_{2k+1}\mu_{2k}t^{2k}\bigg] \ , \qquad N_f=2k+1
\end{equation}
which has been called $\text{HWG}_0$ because it is the highest weight generating function for the topological sector with instanton charge 0. It is exactly the classical HWG for chiral ring of the Higgs branch of $Sp(k)$ with $2k+1$ flavours; furthermore, its interpretation is clear, as it accounts for the mesons (of various orders) and the instanton-anti instanton bound state not at threshold. The fact that
\begin{equation}
    \mu_{2k+1}\mu_{2k}t^{2k}
\end{equation}\label{eq:exampleofinstantoncondensate}
\noindent can be seen as instanton bound state will be clear from the next paragraph, when we will compute the bare instanton. As mentioned previously (see \ref{even flavour condensate at finite coupling}), they can also be interpreted as the natural extension of the higher order mesons, as if the sum in \ref{eq:exampleofinstantoncondensate} was extended by one more term.

\paragraph{Bare instanton}\label{par:bareinstanton}
In order to compute the HWG for the sector of instanton charge $I$, namely $\text{HWG}_I$, it is sufficient to multiply the previous integrand in eq. \ref{eq:2k+1 weyl integration} by $q^I$. In this way, by orthogonality of U(1) character, we will isolate the part of the HWG proportional to $q^I$. We obtain:

\begin{equation}\label{eq:bareinstanton}
    \begin{array}{cc}
        \text{HWG}_{I>0} &=\mu_{2k}^It^{I(k+1)}\cdot \text{HWG}_0  \\
        \\
    \text{HWG}_{I<0}&=\mu_{2k+1}^{|I|}t^{|I|(k+1)}\cdot \text{HWG}_0  \ , \\
    \end{array}
\end{equation}

\noindent where we recognize the bare instanton in the coefficient of the dressing factor. This expression is telling us that the R-charge of the bare instanton of any charge $I$ is $I(k+1)/2$, where $k+1$ is the dual Coxeter number of $Sp(k)$, as expected from \cite{cremonesi2017instanton} and obtained in the case of 4d $\mathcal{N}=1$ in \cite{witten2003chiral}. The $D_{2k+1}$ representation is simply $\mu_{2k}^I$ or $\mu_{2k+1}^I$, which in terms of Dynkin labels translates to:
\begin{equation}
    \begin{array}{cc}
       \mu_{2k}^I:  &  [0,\dots,0,I_{2k},0] \\
       \mu_{2k+1}^I:  &  [0,\dots,0,I_{2k+1}]\\
    \end{array}
\end{equation}
Such representation encodes the degeneracy of fermionic zero modes due to the fundamental strings stretched between the D7-branes and the D0 branes.\\

\noindent We can now interpret eq. \ref{eq:exampleofinstantoncondensate} as an instanton bound state not at threshold, because its R-charge is not the sum of the R-charges of the single instantons. As usual with bound states, the total R-charge is inferior to the sum of the R-charges of the single components. Alternatively, we can interpret such contribution as the $2k$-th antisymmetric product of the fundamental. 

\paragraph{Recovering the initial HWG}
In order to recover the original HWG, in terms of bare instantons and dressing factor, it is necessary to sum over topological sector and multiply by $\text{PE}\big[t^2\big]$:
\[
\bigg(\text{HWG}_0+\text{HWG}_{I>0}+\text{HWG}_{I<0}\bigg)\cdot \text{PE}[t^2]=\text{HWG} \ ,
\]
or explicitely:
\begin{equation}
\bigg(1+\sum_{I=1}^{\infty}q^I\mu_{2k}^It^{I(k+1)}+\sum_{I=1}^{\infty}q^{-I}\mu_{2k+1}^I t^{I(k+1)}\bigg)\cdot \text{HWG}_0\cdot \text{PE}\bigg[t^2\bigg]=\text{HWG} .
\end{equation}
\subsubsection{Bare instanton and dressing factors: cases without symmetry enhancement}
As the discussion in paragraph \ref{par:dressedinstantons} focusses on $N_f=2k+1$, we will now extended the same procedure to the other cases without symmetry enhancement. The results are summarised in Table \ref{tab:dressingandbare for non enhancement}:

\begin{table}[H]
    \centering
    \begin{tabular}{|c|c|c|}\hline
        Number of flavours & Bare instanton & $\text{HWG}_0$ \\ \hline
        & & \\
        \makecell{$N_f\leq 2k+3$\\ $N_f$ odd} &  \makecell{$I>0$: $\mu_{N_f-1}^It^{I(k+1)}$\\ \\ $I<0$: $\mu_{N_f}^{|I|}t^{|I|(k+1)}$} & $\text{PE}\bigg[\sum_{i=1}^{(N_f-1)/2-1}\mu_{2i}t^{2i}+\mu_{N_f}\mu_{N_f-1}t^{N_f-1}\bigg]$\\ 
        & & \\ \hline
         & & \\
          \makecell{$N_f\leq 2k+2$\\ $N_f$ even} & any $I$: $\mu_{N_f}^{|I|}t^{|I|(k+1)}$ & $\text{PE}\bigg[\sum_{i=1}^{N_f/2-1}\mu_{2i}t^{2i}+\mu_{N_f}^2t^{N_f}\bigg]$\\
        & & \\ \hline
    \end{tabular}
    \caption{Bare instantons and dressing factors for colour-flavour combinations without symmetry enhancement.}
    \label{tab:dressingandbare for non enhancement}
\end{table}

\noindent In particular, we notice that for $N_f\leq 2k+1$ the dressing factor is exactly the classical HWG for $Sp(k)$ with $N_f$ flavours. In the case $N_f=2k+2$ the dressing factor can be rewritten by changing $k\to k+1$, finding that it matches with the HWG of one of the two classical cones of $Sp(k+1)$ with $N_f=2k+2$ flavours, namely expression \ref{eq:hwg single cone}. In the case $N_f=2k+3$, the dressing factor is the HWG of the classical chiral ring of $Sp(k+1)$ with same number of flavours.\\
As unified point of view, the variety described by the $\text{HWG}_0$, which we will call \textit{dressing variety}, is always the closure of the maximal nilpotent orbit of $D_{N_f}$ of height two, which has either partition $[2^{N_f}]$, (for even $N_f$) or $[2^{N_f-1},1^2]$, (for odd $N_f$), as reported in table \ref{tab:finite coupling HWG and NOL}.

\subsubsection{Bare instanton and dressing factor: $N_f=2k+4$}
In order to find the bare instanton and dressing in this case, we need to branch the $SU(2)$ representations in terms of the classical U(1) instanton symmetry. The exact algorithm is to transform the HWG into a Hilbert series, branch the $SU(2)$ representation and re-transform into a HWG. However, we proceeded by making an ansatz and matching the coefficients in the perturbative expansions. We find:
\begin{equation}\label{eq:2k+4hwg}
\text{HWG}=\text{PE}\bigg[\sum_{i=1}^{k+1}\mu_{2i}t^{2i}+(1+q^2+q^{-2})t^2+(q+q^{-1})\mu_{2k+4}t^{k+1}\bigg] \ .
\end{equation}
In order to find the dressing factor we need to remember that the instantons and anti instantons appearing in \ref{eq:2k+4hwg} are arranged in $SU(2)$ representations. Explicitely, the character $\chi[n]$ of the representation $[n]$, with highest weight $n$, of $SU(2)$:
\[
\chi[n]=q^n+q^{n-2}+\dots+q^{-(n-2)}+q^{-n}=\sum_{I=-n}^{n}q^{I} \quad \text{in steps of two (i.e. }(n,n-2,n-4)\text{ etc)} 
\]
encodes the presence of an equal number of instantons of positive and negative charge $I$.\\
Therefore the $q$-independent part is found by integrating over $SU(2)$ representations. Similarly, in order to preserve supersymmetry we will need to divide the integrand by the F-constraint, namely PE$\big[[2]t^2\big]$,  which contains the adjoint of SU(2). Explicitely:
\begin{equation}
    \text{HWG}_0=\int \text{d}\mu_{SU(2)}(q) \cdot \text{HWG}\cdot \text{PE}\bigg[-[2]t^2\bigg]=\text{PE}\bigg[\sum_{i=1}^{k+1}\mu_{2i}t^{2i}\bigg] \ ,
\end{equation}
in which we recognize the HWG of the classical chiral ring of $Sp(k+1)$ with $2k+4$ flavours. 
\noindent We can compute the representation of the bare instanton by SU(2) character orthogonality, finding:
\[
\text{HWG}_n= \int \text{d}\mu_{SU(2)}(q)\cdot \chi^*[n]\cdot \text{HWG}\cdot \text{PE}\bigg[-[2]t^2\bigg]=\mu_{2k+4}^{n}t^{n(k+1)}\cdot \text{HWG}_0
\]
It is then straightforward to recover the initial HWG by summing over SU(2) representations:
\begin{equation}
    \sum_{n=0}^{\infty}\chi[n]\cdot \text{HWG}_n \cdot \text{PE}\bigg[[2]t^2\bigg]=\text{HWG} \ .
\end{equation}

\subsubsection{Bare instanton and dressing factor: $N_f=2k+5$}
It was not possible to find the dressing factor and bare instanton for the combination $N_f=2k+5$, as in this case the classical $U(1)$ is embedded into the whole $D_{N_f+1}$ in a way we could not extrapolate.\\
We could however branch the HWG of Table \ref{tab:hwg at infinite coupling} into fugacities of $D_{N_f}$ and charge $q$ of U(1), according to the following Table \ref{eq:2k+5fugacities}:
\begin{equation}\label{eq:2k+5fugacities}
   \begin{array}{|ccc|} \hline
    D_{2k+6} & &  D_{2k+5}\times U(1) \\\hline
    \mu_2 & \to & 1+\mu_1(q^2+q^{-2})+\mu_2 \\
    \mu_4 & \to &  \mu_2 +\mu_3(q^2+q^{-2})+\mu_4 \\
    \vdots & & \vdots \\
    \mu_{2i} & \to & \mu_{2i-2}+\mu_{2i-1}(q^2+q^{-2})+\mu_{2i} \\
    \vdots & & \vdots \\
    \mu_{2k+4} & \to &  \mu_{2k+2}+\mu_{2k+3}(q^2+q^{-2}) \\
    \mu_{2k+6} & \to & q^{-1}\mu_{2k+5}+q\mu_{2k+4} \\\hline
\end{array} 
\end{equation}
The resulting HWG is:
\begin{equation}
\text{HWG}=\text{PE}\bigg[\sum_{i=1}^{k+1}\mu_{2i}t^{2i}(1+t^2)+\sum_{i=1}^{k+2}\mu_{2i-1}t^{2i}(q^2+q^{-2})+t^2+t^4+(q^{-1}\mu_{2k+5}+q\mu_{2k+4})(1+t^{2})t^{k+1}\bigg]  \ ,
\end{equation}
which can be recast in:
\begin{equation}
    \begin{array}{ll}
       \text{HWG}=  &\text{PE}\bigg[\sum_{i=1}^{k+1}\mu_{2i}t^{2i}+t^2+(q^{-1}\mu_{2k+5}+q\mu_{2k+4})t^{k+1} \bigg] \times  \\
         & \times \text{PE}\bigg[\sum_{i=1}^{k+1}\mu_{2i}t^{2i+2}+t^4+(q^{-1}\mu_{2k+5}+q\mu_{2k+4})t^{k+3} \bigg]\times  \\
        & \times \text{PE}\bigg[\sum_{i=0}^{k+1}\mu_{2i+1}t^{2i+2}(q^2+q^{-2})\bigg] \ . \\
    \end{array}\label{decomposed HWg}
\end{equation}
Although unable to further simplify these expressions, let us point out that \ref{decomposed HWg} has the non-trivial property of being itself the PE of a polynomial.

\subsection{General expressions for bare instantons and dressing factor}
\paragraph{Bare instanton}
We can write a unique expression for the bare instanton at charge $I$ (modulo $N_f$ oddness), for any $N_f$ and $k$:
\begin{table}[H]
    \centering
    \begin{tabular}{|c|c|c|c|}\hline
       Even/odd  &  Bare instanton at charge $I$ & $D_{N_f}$ rep. & R-charge\\ \hline
       & & &  \\
       $N_f$ even  &  $\mu_{N_f}^{|I|}t^{|I|(k+1)}$ & $[0,\dots,0,|I|]$ & $\dfrac{|I|}{2}(k+1)$\\
       & & & \\ \hline
       & & & \\
       $N_f$ odd & \makecell{$I>0$: \quad $\mu_{N_f-1}^{I}t^{I(k+1)}$\\ \\$I<0$: \qquad  $\mu_{N_f}^{|I|}t^{|I|(k+1)}$ } & \makecell{$[0,\dots,0,I,0]$ \\ \\$[0,\dots,0,|I|]$} & $\dfrac{|I|}{2}(k+1)$\\
       & & &  \\ \hline
    \end{tabular}
    \caption{Quantum numbers of bare instanton for any $N_f$ and $k$ (except for $N_f=2k+5$). The dual Coxeter number of $Sp(k)$, which is $k+1$, appears in the R-charge.}
    \label{tab:my_label}
\end{table}
\paragraph{Dressing factor}
For what concerns the dressing factor, an important difference has to be made between $N_f\leq 2k+1$ and $N_f>2k+1$. Indeed, in the former case the instanton-bound state is not at threshold, coherently with the HWG of the classical Higgs branch. In the latter case, the bound state is at threshold. By performing a $k\to k+1$, the latter bound state can be seen as the HWG of $Sp(k+1)$ with $N_f$ flavours. Let us summarize this result in Table \ref{tab:dressingfactorsingeneral}.

\begin{table}[H]
    \centering
    \begin{tabular}{|c||c|c|c|} \hline
       Range   & Dressing factor & Classical interpretation & $\bar{\mathcal{O}}_D$\\ \hline \hline
       & & & \\
    $N_f=2k+4$  & $\text{PE}\bigg[\sum_{i=1}^{k+1}\mu_{2i}t^{2i}\bigg]$ & \makecell{$Sp(k+1)$ with\\ $2k+4$ flavours} & $\bar{\mathcal{O}}_D^{[2^{2k+2},1^{4}]}$  \\
    & & & \\ \hline
    & & & \\
   \makecell{$N_f\leq 2k+3$\\ $N_f$ odd} & \makecell{$\text{PE}\bigg[\sum_{i=1}^{(N_f-3)/2}\mu_{2i}t^{2i}+$\\$+\mu_{N_f}\mu_{N_f-1}t^{N_f-1}\bigg]$} & \makecell{$Sp\bigg(\dfrac{N_f-1}{2}\bigg)$ with\\ $N_f\leq 2k+3$ flavours} & $\bar{\mathcal{O}}_D^{[2^{N_f-1},1^2]}$ \\
    & & & \\ \hline
    & & & \\
    \makecell{$N_f\leq 2k+2$\\ $N_f$ even} & $\text{PE}\bigg[\sum_{i=1}^{N_f/2-1}\mu_{2i}t^{2i}+\mu_{N_f}^2t^{N_f}\bigg]$ & \makecell{Single cone of $Sp(N_f/2)$ with\\$N_f\leq 2k+2$ flavours} & $\bar{\mathcal{O}}_D^{[2^{N_f}]}$ \\
    & & & \\ \hline
    \end{tabular}
    \caption{Dressing factor for any $N_f$ and $k$ (except for $N_f=2k+5$).}
    \label{tab:dressingfactorsingeneral}
\end{table}

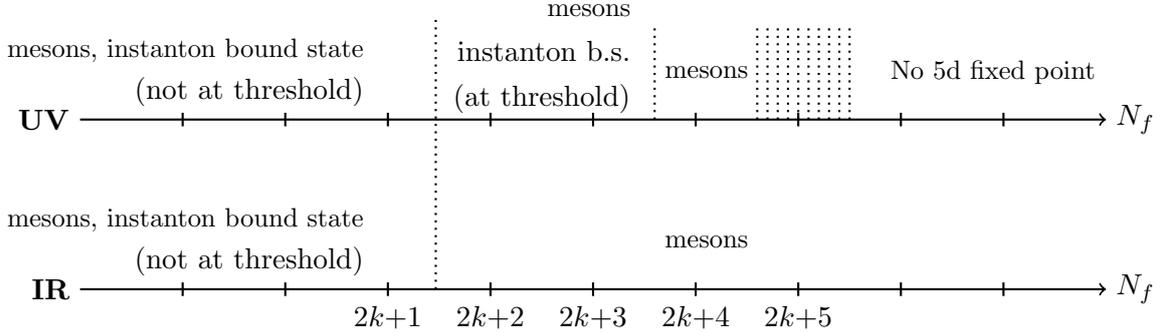
\begin{figure}[H]
    \centering
        \begin{tikzpicture}[scale=0.9]

\draw[->, thick] (0,2.5) -- (15,2.5);  
\node[left] at (0,2.5) {\textbf{UV}};
\node[right] at (15,2.5) {\(N_f\)};
\foreach \i in {1,...,9} {
  \pgfmathsetmacro\x{1.5*\i}
  \draw[thick] (\x,2.6) -- (\x,2.4);
}

\draw[->, thick] (0,0) -- (15,0);
\node[left] at (0,0) {\textbf{IR}};
\node[right] at (15,0) {\(N_f\)};
\foreach \i in {1,...,9} {
  \pgfmathsetmacro\x{1.5*\i}
  \draw[thick] (\x,0.1) -- (\x,-0.1);
}

\node[below] at (4.5,-0.1) {\(2k{+}1\)};
\node[below] at (6.0,-0.1) {\(2k{+}2\)};
\node[below] at (7.5,-0.1) {\(2k{+}3\)};
\node[below] at (9.0,-0.1) {\(2k{+}4\)};
\node[below] at (10.5,-0.1) {\(2k{+}5\)};

\draw[dotted, thick] (5.2,0) -- (5.2,4.0);
\draw[dotted, thick] (10.5,2.5) -- (10.5,3.9);
\draw[dotted, thick] (8.4,2.5) -- (8.4,3.9);
\draw[dotted, thick] (9.9,2.5) -- (9.9,3.9);
\draw[dotted, thick] (10.05,2.5) -- (10.05,3.9);
\draw[dotted, thick] (10.2,2.5) -- (10.2,3.9);
\draw[dotted, thick] (10.35,2.5) -- (10.35,3.9);
\draw[dotted, thick] (10.65,2.5) -- (10.65,3.9);
\draw[dotted, thick] (10.8,2.5) -- (10.8,3.9);
\draw[dotted, thick] (10.95,2.5) -- (10.95,3.9);
\draw[dotted, thick] (11.1,2.5) -- (11.1,3.9);
\draw[dotted, thick] (11.25,2.5) -- (11.25,3.9);

\node[align=right, anchor=east] at (4.3,0.7) {\small mesons, instanton bound state\\[1pt] (not at threshold)};
\node[align=left, anchor=west] at (8.4,0.7) {\small mesons};

\node[align=right, anchor=east] at (4.3,3.2) {\small mesons, instanton bound state\\[1pt] (not at threshold)};
\node[align=left, anchor=west] at (11.7,3.2) {\small No 5d fixed point};

\node[align=right, anchor=east] at (8.2,3.4) {\small mesons\\[2pt] instanton b.s.\\[2pt] (at threshold)};
\node[align=left, anchor=west] at (8.4,3.2) {\small mesons};

\end{tikzpicture}
    \caption{Comparison between dressing factor in the UV and HWG in the IR, for Sp(k) with $N_f$ flavours.}
    \label{fig:dressingfactordiagram}
\end{figure}

\noindent We will now present a diagrammatical depiction of the argument of the dressing factor, from IR to UV, in Figure \ref{fig:dressingfactordiagram}.

\noindent As reported in Figure \ref{fig:dressingfactordiagram}, the HWG depends in the UV on the instanton bound state. However its R-charge changes across $N_f=2k+1$: for lower number of flavours, the bound state is not at threshold, while it is at threshold for higher number of flavours. The reason of this transition has to be seen in the competition of the first and second constraints in \ref{eq:Josephrelations}. At fixed $k$, for $N_f \leq 2k+1$ the $M^2=0$ relation is stronger, while for $N_f\geq 2k+2$ the $\text{rank}M\leq 2k$ relation wins.

\subsection{Dressing varieties from the magnetic quiver}
The magnetic quivers for the dressing varieties can be found by ungauging a U(1) node, or an SU(2) node in the case of $N_f=2k+4$, from the infinite coupling magnetic quiver. The result of this ungauging procedure is presented in Table \ref{tab:dressingvarieties}:

\begin{longtable}{|>{\centering\arraybackslash}m{2.5cm}|l|} \hline
        Flavours & Quiver  \\ \hline
     $N_f=2k+4$  & \begin{tikzpicture}[
  scale=1, every node/.style={draw, circle, minimum size=5pt, inner sep=0pt},
  label distance=2mm
]

  \node (A1) at (0,0) {};
  \node (A4) at (2,0) {};
  \node (A5) at (3.5,0) {};

  \node[draw=none, below=1mm of A1] {1};
  \node[draw=none, below=-3mm of A4] {$2k+2$};
  \node[draw=none, above=-1mm of A5] {$k+2$};
  
  \draw[dotted] (A1) -- (A4);

  \node (Top) at (2,1) {};
  \node[draw=none, above=-1mm of Top] {$k+1$};
  \draw (A4) -- (Top);

  \node[draw=none] (NE) at (5,0) {};
  
  \node[draw, rectangle, minimum size=5pt, inner sep=0pt, label=below:2] at (5,0) {};

  \draw (A4) -- (A5);
  \draw (A5) -- (NE);

\end{tikzpicture} \\  \hline \vspace{1cm}
$N_f=2k+3$ & 
 \begin{tikzpicture}[
  scale=1, every node/.style={draw, circle, minimum size=5pt, inner sep=0pt},
  label distance=2mm
]

  \node (A1) at (0,-1) {};
  \node (A4) at (2,-1) {};
  \node (A5) at (3.5,-1) {};
  \node[draw=none] (A6) at (2,1) {};

  \node[draw=none, below=1mm of A1] {1};
  \node[draw=none, below=-3mm of A4] {$2k+1$};
  \node[draw=none, below=-2mm of A5] {$k+1$};
  
  \draw[dotted] (A1) -- (A4);

  \node (Top) at (2,0) {};
  \node[draw=none, left=0mm of Top] {$k+1$};
  \draw (A4) -- (Top);
\node[draw, rectangle, minimum size=5pt, inner sep=0pt] at (2,1) {};
  \node[draw=none, above=1mm of A6] {$1$};
  \draw (Top) -- (A6);

  \node[draw=none] (NE) at (5,-1) {};
  \node[draw=none, below=1mm of NE] {1};

  \node[draw, rectangle, minimum size=5pt, inner sep=0pt] at (5,-1) {};
  \draw (A4) -- (A5);
  \draw (A5) -- (NE);
\end{tikzpicture}
\\ \hline
$N_f=2k+2$  & \begin{tikzpicture}[
  scale=1, every node/.style={draw, circle, minimum size=5pt, inner sep=0pt},
  label distance=2mm
]

  \node (A1) at (0,-1) {};
  \node (A4) at (2,-1) {};
  \node (A5) at (3.5,-1) {};

  \node[draw=none, below=1mm of A1] {1};
  \node[draw=none, below=0mm of A4] {$2k$};
  \node[draw=none, below=-2mm of A5] {$k+1$};
  
  \draw[dotted] (A1) -- (A4);

  \node (Top) at (2,0) {};
  \node[draw=none, above=0mm of Top] {$k$};
  \draw (A4) -- (Top);
\node[draw=none, above=5mm of A6] {};
 \draw (A4) -- (A5);

  \node[draw, rectangle, minimum size=5pt, inner sep=0pt] (NE) at (5,-1) {}; 
 
  \node[draw=none, below=1mm of NE] {2};
  \draw (A5) -- (NE);
  
\end{tikzpicture} \\ \hline
\caption{Magnetic quivers for the dressing varieties. For $N_f\leq 2k+1$, the quivers are the same as the one in Table \ref{tab:magnetic quivers classical theory}, as there is no dependence on $k$.}  \label{tab:dressingvarieties}\\
\end{longtable}
\noindent Finding the dressing variety by means of ungauging a U(1) gauge group from the magnetic quiver could be a guiding principle whenever the HWG is computationally hard to deal with.

\section{Conclusions}
We have explained how the chiral ring of the Higgs branch at infinite coupling, for the theories at hand, can be simply related to the one at finite coupling as in:
\begin{equation}
    \text{HWG}^{\text{UV}} = \text{tr}_q[I_q]\cdot \text{PE}[t^2]\cdot \text{HWG}^{\text{IR}}
\end{equation}
where $\text{HWG}^{\text{IR}}$ is the HWG for the chiral ring of (a single cone of) $Sp(k+1)$ with same amount of flavours.\\
Furthermore, we have also shown the $D_{N_f}$ representation and R-charge at which the instanton of generic charge arises. As argued from the braneweb construction, such representation is the consequence of the degeneracy of zero modes of the strings stretched between the D1 and the D7 branes. This extends the previous results in the literature, where only for the charge 1 or charge 2 instantons one could specify the $D_{N_f}$ representation or R-charge (usually by computation of the superconformal index \cite{zafrir2015instanton, kim20125}).\\

\noindent Due to the well known \cite{bourget2020magnetic} duality of:
\begin{equation}
    \begin{array}{ccc}
        Sp(k) \quad \text{with} \; N_f \;\text{flavours} & \longleftrightarrow & SU(k+1)_{\pm(k+3-N_f/2)} \quad \text{with}  \; N_f \;\text{flavours}\\
         & 
    \end{array}
\end{equation}
the results of this paper can be extended to $SU$ theories with suitable CS level. A possible extension of this work is to other CS levels or to different classical Lie groups altogether, although it is foreseeable that the presence of a baryonic symmetry would make harder to disentangle the infrared U(1) instanton symmetry. More general representation of the matter content can also be considered.\\

\noindent As far as future directions of research are concerned, it could be interesting to study more closely the instanton anti-instanton bound state, perhaps in the spirit of \cite{lambert2015instanton}, trying in particular to determine whether this interpretation is more informative than the one describing $\mu_{2k}\mu_{2k+1}$ as the symmetric product of adjoint representations.\\

\noindent More generally, the description of Higgs branches as moduli space of dressed instantons would be useful whenever those branches could not be encoded as moduli space of dressed monopoles. It would be interesting to understand to what extent they offer additional insights or a convenient classification.\\
\vspace{1cm}

\section*{Acknowledgments}
E.VdD. would like to thank Guhesh Kumaran, Sam Bennett and Rudolph Kalveks for interesting discussions. The work of A.H. and E.VdD. is partially supported by STFC Consolidated Grant ST/X000575/1.

\clearpage

\bibliographystyle{JHEP}
\bibliography{bibli.bib}

@article{lambert2015instanton,
  title={Instanton operators in five-dimensional gauge theories},
  author={Lambert, Neil and Papageorgakis, C and Schmidt-Sommerfeld, M},
  journal={Journal of High Energy Physics},
  volume={2015},
  number={3},
  pages={1--16},
  year={2015},
  publisher={Springer},
  eprint={1412.2789},
  archivePrefix={arXiv},
  primaryClass={hep-th}
}

@article{bourget2020higgs,
  title={The Higgs mechanism—Hasse diagrams for symplectic singularities},
  author={Bourget, Antoine and Cabrera, Santiago and Grimminger, Julius F and Hanany, Amihay and Sperling, Marcus and Zajac, Anton and Zhong, Zhenghao},
  journal={Journal of High Energy Physics},
  volume={2020},
  number={1},
  pages={1--67},
  year={2020},
  publisher={Springer},
  eprint={1908.04245},
  archivePrefix={arXiv},
  primaryClass={hep-th}
}

@article{bergman2014discrete,
  title={Discrete $\theta$ and the 5d superconformal index},
  author={Bergman, Oren and Rodr{\'\i}guez-G{\'o}mez, Diego and Zafrir, Gabi},
  journal={Journal of High Energy Physics},
  volume={2014},
  number={1},
  pages={1--14},
  year={2014},
  publisher={Springer},
  eprint={1310.2150},
  archivePrefix={arXiv},
  primaryClass={hep-th}
}

@article{bourget2020brane,
  title={Brane webs and magnetic quivers for SQCD},
  author={Bourget, Antoine and Cabrera, Santiago and Grimminger, Julius F and Hanany, Amihay and Zhong, Zhenghao},
  journal={Journal of High Energy Physics},
  volume={2020},
  number={3},
  pages={1--58},
  year={2020},
  publisher={Springer},
  eprint={1909.00667},
  archivePrefix={arXiv},
  primaryClass={hep-th}
}

@article{bourget2020magnetic,
  title={Magnetic quivers from brane webs with O5 planes},
  author={Bourget, Antoine and Grimminger, Julius F and Hanany, Amihay and Sperling, Marcus and Zhong, Zhenghao},
  journal={Journal of High Energy Physics},
  volume={2020},
  number={7},
  pages={1--82},
  year={2020},
  publisher={Springer},
  eprint={2004.04082},
  archivePrefix={arXiv},
  primaryClass={hep-th}
}

@article{benvenuti2007counting,
  title={Counting BPS operators in gauge theories: quivers, syzygies and plethystics},
  author={Benvenuti, Sergio and Feng, Bo and Hanany, Amihay and He, Yang-Hui},
  journal={Journal of High Energy Physics},
  volume={2007},
  number={11},
  pages={050},
  year={2007},
  publisher={IOP Publishing},
  eprint={hep-th/0608050},
  archivePrefix={arXiv},
  primaryClass={hep-th}
}

@article{kim2015tao,
  title={Tao probing the end of the world},
  author={Kim, Sung-Soo and Taki, Masato and Yagi, Futoshi},
  journal={Progress of Theoretical and Experimental Physics},
  volume={2015},
  number={8},
  pages={083B02},
  year={2015},
  publisher={Oxford University Press},
  eprint={1504.03672},
  archivePrefix={arXiv},
  primaryClass={hep-th}
}

@article{intriligator1997five,
  title={Five-dimensional supersymmetric gauge theories and degenerations of Calabi-Yau spaces},
  author={Intriligator, Kenneth and Morrison, David R and Seiberg, Nathan},
  journal={Nuclear Physics B},
  volume={497},
  number={1-2},
  pages={56--100},
  year={1997},
  publisher={Elsevier},
  eprint={hep-th/9702198},
  archivePrefix={arXiv},
  primaryClass={hep-th}
}

@article{morrison1997extremal,
  title={Extremal transitions and five-dimensional supersymmetric field theories},
  author={Morrison, David R and Seiberg, Nathan},
  journal={Nuclear Physics B},
  volume={483},
  number={1-2},
  pages={229--247},
  year={1997},
  publisher={Elsevier},
  eprint={hep-th/9609070},
  archivePrefix={arXiv},
  primaryClass={hep-th}
}

@article{cabrera2018quiver,
  title={Quiver subtractions},
  author={Cabrera, Santiago and Hanany, Amihay},
  journal={Journal of High Energy Physics},
  volume={2018},
  number={9},
  pages={1--21},
  year={2018},
  publisher={Springer},
  eprint={1803.11205},
  archivePrefix={arXiv},
  primaryClass={hep-th}
}

@article{hanany2014highest,
  title={Highest weight generating functions for Hilbert series},
  author={Hanany, Amihay and Kalveks, Rudolph},
  journal={Journal of High Energy Physics},
  volume={2014},
  number={10},
  pages={1--68},
  year={2014},
  publisher={Springer},
  eprint={1408.4690},
  archivePrefix={arXiv},
  primaryClass={hep-th}
}

@article{cremonesi2014monopole,
  title={Monopole operators and Hilbert series of Coulomb branches of 3d N= 4 gauge theories},
  author={Cremonesi, Stefano and Hanany, Amihay and Zaffaroni, Alberto},
  journal={Journal of High Energy Physics},
  volume={2014},
  number={1},
  pages={1--34},
  year={2014},
  publisher={Springer},
  eprint={1309.2657},
  archivePrefix={arXiv},
  primaryClass={hep-th}
}

@article{aharony1998webs,
  title={Webs of (p, q) 5-branes, five dimensional field theories and grid diagrams},
  author={Aharony, Ofer and Hanany, Amihay and Kol, Barak},
  journal={Journal of High Energy Physics},
  volume={1998},
  number={01},
  pages={002},
  year={1998},
  publisher={IOP Publishing},
  eprint={hep-th/9710116},
  archivePrefix={arXiv},
  primaryClass={hep-th}
}

@article{cabrera2019tropical,
  title={Tropical geometry and five dimensional Higgs branches at infinite coupling},
  author={Cabrera, Santiago and Hanany, Amihay and Yagi, Futoshi},
  journal={Journal of High Energy Physics},
  volume={2019},
  number={1},
  pages={1--50},
  year={2019},
  publisher={Springer},
  eprint={1810.01379},
  archivePrefix={arXiv},
  primaryClass={hep-th}
}

@article{aharony1997branes,
  title={Branes, superpotentials and superconformal fixed points},
  author={Aharony, Ofer and Hanany, Amihay},
  journal={Nuclear Physics B},
  volume={504},
  number={1-2},
  pages={239--271},
  year={1997},
  publisher={Elsevier},
  eprint={hep-th/9704170},
  archivePrefix={arXiv},
  primaryClass={hep-th}
}

@article{ferlito2016tale,
  title={A tale of two cones: the Higgs Branch of Sp (n) theories with 2n flavours},
  author={Ferlito, Giulia and Hanany, Amihay},
  journal={arXiv preprint arXiv:1609.06724},
  year={2016},
  eprint={1609.06724},
  archivePrefix={arXiv},
  primaryClass={hep-th}
}

@article{bergman20155d,
  title={5d fixed points from brane webs and O7-planes},
  author={Bergman, Oren and Zafrir, Gabi},
  journal={Journal of High Energy Physics},
  volume={2015},
  number={12},
  pages={1--30},
  year={2015},
  publisher={Springer},
  eprint={1507.03860},
  archivePrefix={arXiv},
  primaryClass={hep-th}
}

@article{Bourget:2023cgs,
    author = "Bourget, Antoine and Grimminger, Julius F. and Hanany, Amihay and Kalveks, Rudolph and Sperling, Marcus and Zhong, Zhenghao",
    title = "{A tale of N cones}",
    eprint = "2303.16939",
    archivePrefix = "arXiv",
    primaryClass = "hep-th",
    doi = "10.1007/JHEP09(2023)073",
    journal = "JHEP",
    volume = "09",
    pages = "073",
    year = "2023"
}

@article{Hanany:2016gbz,
    author = "Hanany, Amihay and Kalveks, Rudolph",
    title = "{Quiver Theories for Moduli Spaces of Classical Group Nilpotent Orbits}",
    eprint = "1601.04020",
    archivePrefix = "arXiv",
    primaryClass = "hep-th",
    doi = "10.1007/JHEP06(2016)130",
    journal = "JHEP",
    volume = "06",
    pages = "130",
    year = "2016"
}

@article{witten2003chiral,
  title={Chiral ring of Sp (N) and SO (N) supersymmetric gauge theory in four dimensions},
  author={Witten, Edward},
  journal={Chinese Annals of Mathematics},
  volume={24},
  number={04},
  pages={403--414},
  year={2003},
  publisher={World Scientific},
  eprint={hep-th/0302194},
  archivePrefix={arXiv},
  primaryClass={hep-th}
}

@article{seiberg1996five,
  title={Five dimensional SUSY field theories, non-trivial fixed points and string dynamics},
  author={Seiberg, Nathan},
  journal={Physics Letters B},
  volume={388},
  number={4},
  pages={753--760},
  year={1996},
  publisher={Elsevier},
  eprint={hep-th/9608111},
  archivePrefix={arXiv},
  primaryClass={hep-th}
}

@article{zafrir2015instanton,
  title={Instanton operators and symmetry enhancement in 5d supersymmetric USp, SO and exceptional gauge theories},
  author={Zafrir, Gabi},
  journal={Journal of High Energy Physics},
  volume={2015},
  number={7},
  pages={1--34},
  year={2015},
  publisher={Springer},
  eprint={1503.08136},
  archivePrefix={arXiv},
  primaryClass={hep-th}
}

@article{kim20125,
  title={5-dim superconformal index with enhanced E n global symmetry},
  author={Kim, Hee-Cheol and Kim, Sung-Soo and Lee, Kimyeong},
  journal={Journal of High Energy Physics},
  volume={2012},
  number={10},
  pages={1--66},
  year={2012},
  publisher={Springer},
  eprint={1206.6781},
  archivePrefix={arXiv},
  primaryClass={hep-th}
}

@article{cremonesi2017instanton,
  title={Instanton operators and the Higgs branch at infinite coupling},
  author={Cremonesi, Stefano and Ferlito, Giulia and Hanany, Amihay and Mekareeya, Noppadol},
  journal={Journal of High Energy Physics},
  volume={2017},
  number={4},
  pages={1--43},
  year={2017},
  publisher={Springer},
  eprint={1505.06302},
  archivePrefix={arXiv},
  primaryClass={hep-th}
}

\end{document}